\documentclass[aps,preprint,showpacs,floatfix]{revtex4}%
\usepackage{amsfonts}
\usepackage{amsmath}
\usepackage{amssymb}
\usepackage{graphicx}
\usepackage{graphicx}%
\begin{document}
\title{Frequency crystal}
\author{R. N. Shakhmuratov}
\affiliation{Zavoisky Physical-Technical Institute, FRC Kazan Scientific Center of RAS,
Kazan 420029, Russia and Kazan Federal University, 18 Kremlyovskaya Street,
Kazan 420008, Russia}
\date{{ \today}}

\begin{abstract}
Pulse filtering through a medium with infinite periodic structure in
transmission spectrum is analyzed. Two types of filters are considered. The
first, named harmonic frequency crystal (HFC), is the filter whose widths of
transmitting and absorbing windows are equal. The second, named anharmonic
frequency crystal (AHFC), has narrow absorption peaks separated by wide
transmission windows. AHFC of moderate optical thickness demonstrates
properties quite similar to those of high finesse atomic frequency comb (AFC)
with limited number of the absorption peaks, which produce a few pulses from a
short input pulse. On the contrary, HFC transforms the input pulse into a
train of pulses whose maximum amplitudes follow a wide bell-shaped envelope.
HFC allows to find an exact universal analytical solution, which describes
transformation of a broadband pulse into a train of short pulses, slow light
propagation for a pulse whose spectral width fits one of the transparency
windows of the crystal, and absorption typical for a single line absorber if
the pulse spectrum falls into one of the absorption peaks. Potential
applications of HFC are discussed.

\end{abstract}
\maketitle

\section{Introduction}

Infinite periodical structures in space, time and frequency are models, which
allow simple analytical solutions. Ideal crystals infinite in space are nice
models to predict electron, phonon and thermodynamical properties of real
crystals. A time crystal or space-time crystal, which is a structure that
repeats in time, as well as in space, recently was proposed to create objects
with unusual thermodynamical properties and never reaching thermal equilibrium
\cite{Wilczek,Sacha,Khemani,Else,Sondhi,Yao,Lukin}. Frequency crystals with a
periodic structure in absorption spectrum of electromagnetic field were
proposed as a quantum memory for repeaters
\cite{Gisin2008,Gisin2009,Lauro,Bonarota,Chaneliere,Bonarota2012}. These
crystals were not ideal since they have finite length in frequency domain with
decreasing depths of the transparency windows at the edges of the spectrum.
They were created in inhomogeneously broadened absorption spectrum of crystals
with rare-earth-metal impurity ions by spectral hole burning technique.
Therefore, these frequency crystals were named atomic frequency combs (AFC).

In this paper, the ideal crystals are considered with infinite periodic
structure of transparent holes in the absorption spectrum. The sequence of
transmission and absorption windows follow harmonic law and can be described
by sine function. Therefore, the absorber with such a structure can be named
harmonic frequency crystal (HFC). The harmonicity allows to derive an exact
solution with a very simple structure, which discloses physical properties of
the generated pulses at the exit of HFC. In contrast to AFC, HFC generates a
series of pulses whose maximum amplitudes build a bell-shaped envelope with
much longer duration than the interval $T$ between pulses, which is inverse
value of the frequency period of the comb. Numerical simulations confirm
validity of the exact solution. It would point out that infinite spectrum of
HFC is necessary only for derivation of the solution, while the result
describes perfectly the case if one truncates in numerical simulations the
spectrum of HFC to finite boundaries where the pulse spectrum has noticeable power.

The exact solution (ES) is universal and describes not only the case when the
spectrum of the incident light pulse covers many absorption peaks, but also
slow propagation of the pulse when its spectrum is fully inside one of the
transparency windows of the comb. ES describes also a case if the pulse
spectrum is confined within one of the absorption peaks. Approximate
analytical and numerical simulations confirm validity of ES for both cases.

Results for HFC are compared with numerical simulations and analytical
calculations for anharmonic frequency combs (AHFC). Two examples are
considered. In one the transmission windows follow the law $\sin^{10}(\pi
\nu/2\nu_{0})$, where $\nu=\omega_{c}-\omega$ is the frequency difference
between the central frequency of the light pulse $\omega_{c}$ and its spectral
component $\omega$, and $\nu_{0}$ is the distance from the center of the
transmission window and the nearest absorption peak. The second example is a
sequence of Lorenzian absorption peaks whose width could be much smaller than
the distance between them. Such a structure is equivalent to AFC with high
finesse \cite{Gisin2008,Gisin2009,Lauro,Bonarota,Chaneliere,Bonarota2012}.

Potentially harmonic frequency crystal could be applied to generate a long
train of short coherent pulses with long net-coherence length, which are quite
important for creating time-bin qubits of high dimension from a faint laser
pulse. Recently, the transformation of a long pulse with a comb spectrum into
a train of coherent short pulses by filtering through a single-narrow-line
absorber was proposed and experimentally implemented with single $\gamma
$-photons \cite{Vagizov,Shakhmuratov15,Shakhmuratov17}. In this paper, the
transformation of a short pulse into a long train of coherent pulses by
filtering through a comb structure is proposed. Another promising application
is optical detection of ultrasound in biological tissues for
ultrasound-modulated optical tomography. It was shown that AFC allows high
discrimination between the sidebands and the carrier \cite{McAuslan,Horiuchi}.
One could expect that HFC could appreciably increase this discrimination and
the signal to noise ratio.

The paper is organized as follows. In Sec. II the definition of harmonic
frequency crystal is introduced. With the help of Kramers-Kronig relation the
transmission function of HFC is derived. In Sec. II exact solution for the
pulse propagating in HFC is obtained. The properties of ES are discussed in
Sec. IV. The application of ES \ for the description of slow light propagation
in a medium with a transparency window is demonstrated in Sec. V. ES
application for the description of the resonant absorption of the pulse whose
spectrum is completely inside one of the absorption peaks of HFC is discussed
in Sec. VI. The properties of the anharmonic frequency combs are discussed in
Secs. VII and VIII. The results are summarized in Sec. IX.

\section{Harmonic frequency crystal}

We start with a very general consideration of the propagation of a radiation
pulse through a medium whose response is described by
\begin{equation}
\mathbf{P}=\varepsilon_{0}\chi\mathbf{E}, \label{Eq1}%
\end{equation}
where $\mathbf{E}$ is an electric field, $\mathbf{P}$ is polarization density,
$\varepsilon_{0}$ is the electric permittivity of free space (below we set
$\varepsilon_{0}=1$ for simplicity) and $\chi$ is the electric susceptibility,
which is
\begin{equation}
\chi=\chi^{\prime}+i\chi^{\prime\prime}. \label{Eq2}%
\end{equation}
Imaginary part of susceptibility $\chi^{\prime\prime}$ describes the field
absorption. We suppose that $\chi^{\prime\prime}$ is a harmonic periodic
function%
\begin{equation}
\chi_{\mathrm{h}}^{\prime\prime}(\nu)=\frac{\chi_{0}}{2}\left[  1-\cos\left(
\frac{\pi\nu}{\nu_{0}}\right)  \right]  , \label{Eq3}%
\end{equation}
which is infinite in frequency space. Here, index $\mathrm{h}$ means harmonic,
$\chi_{0}$ is the value, which depends on interaction constant with the medium
and its density, $\nu=\omega_{c}-\omega$ is the frequency difference between
the central frequency of the light pulse $\omega_{c}$ and its spectral
component $\omega$.

Equation (\ref{Eq3}) describes a periodic structure of transmission windows
and absorption peaks with a distance between centers of the transparency
windows and nearest absorption peaks equal to $\nu_{0}$. The widths of the
peaks and windows are equal the same value $\nu_{0}$. Meanwhile, finesse $F$
of the comb structure, defined in Ref. \cite{Gisin2009} as the ratio of the
distance between the absorption peaks $\Delta_{a}$ and the width of the
individual peak $w_{a}$, i.e., $F=\Delta_{a}/w_{a}$, is equal $2$ for the
harmonic frequency crystal since its parameters are $\Delta_{a}=2\nu_{0}$ and
$w_{a}=\nu_{0}$. Transmission windows of HFC are ideal, i.e., there is no
absorption at the window center.

To satisfy causality the dispersion relations or more generally Kramers-Kronig
relations must be fulfilled that describe the frequency dependence of the wave
propagation and attenuation. According to them the real part of the
susceptibility, responsible for a group velocity dispersion, satisfies one of
the Kramers-Kronig relations
\begin{equation}
\chi^{\prime}(\nu)=\frac{1}{\pi}\mathcal{P}%
{\displaystyle\int\nolimits_{-\infty}^{\infty}}
\frac{\chi^{\prime\prime}(\nu^{\prime})}{\nu^{\prime}-\nu}d\nu^{\prime},
\label{Eq4}%
\end{equation}
where $\mathcal{P}$ denotes the Cauchy principal value. Calculating the
integral, we obtain
\begin{equation}
\chi_{\mathrm{h}}^{\prime}(\nu)=\frac{\chi_{0}}{2}\sin\left(  \frac{\pi\nu
}{\nu_{0}}\right)  . \label{Eq5}%
\end{equation}
Unidirectional wave equation, describing the propagation of the plane-wave
field $E(z,t)=E_{0}(z,t)\exp(-i\omega_{c}t+ikz)$ along axis $\mathbf{z}$, is
(see, for example, Ref. \cite{Crisp})%
\begin{equation}
\left(  \frac{\partial}{\partial z}+\frac{n}{c}\frac{\partial}{\partial
t}\right)  E_{0}(z,t)=i\frac{2\pi\omega_{c}}{nc}P_{0}(z,t), \label{Eq6}%
\end{equation}
where $E_{0}(z,t)$ is the pulse envelope, $k$ is the wave number,
$P(z,t)=P_{0}(z,t)\exp(-i\omega_{c}t+ikz)$ is the polarization induced in the
medium, and $n$ index of refraction (below we set $n=1$ for simplicity).

By the Fourier transform
\begin{equation}
F(\nu)=\int_{-\infty}^{+\infty}f(t)e^{i\nu t}dt,\label{Eq7}%
\end{equation}
Eq. (\ref{Eq6}) is reduced to one-dimensional differential equation, whose
solution is%
\begin{equation}
E_{0}(z,\nu)=E_{0}(0,\nu)\exp\left\{  i\nu\frac{z}{c}-\frac{\alpha_{p}z}%
{2\chi_{0}}\left[  \chi_{\mathrm{h}}^{\prime\prime}(\nu)-i\chi_{\mathrm{h}%
}^{\prime}(\nu)\right]  \right\}  ,\label{Eq8}%
\end{equation}
where $E_{0}(0,\nu)=E_{in}(\nu)$ is a spectral component of the field incident
to the frequency crystal, $\alpha_{p}$ is the Beer's law attenuation
coefficient describing absorption of a monochromatic field tuned in resonance
with one of the absorption peaks.

For the harmonic frequency crystal of physical length $l$ we have%
\begin{equation}
E_{0}(l,\nu)=E_{0}(0,\nu)\exp\left\{  i\nu\frac{l}{c}-\frac{d_{p}}{4}%
+\frac{d_{p}}{4}e^{i\pi\nu/\nu_{0}}\right\}  , \label{Eq9}%
\end{equation}
where $d_{p}=\alpha_{p}l$ is the absorption depth (optical thickness) of the
medium for the monochromatic radiation field tuned in resonance with one of
the absorption peaks. Below, for simplicity, we neglect small value $i\nu l/c$
in the exponent, which is responsible for small delay due to the pulse
traveling the distance $l$ with a speed of light $c$ ($t_{d}=l/c$).

\section{Exact solution}

Infinite periodical structures in the absorption and dispersion, Eqs.
(\ref{Eq3}), (\ref{Eq5}), allow to find an exact solution (ES), which simply
follows from the Taylor series
\begin{equation}
e^{(d_{p}/4)e^{i\pi\nu/\nu_{0}}}=\sum_{k=0}^{+\infty}\frac{(d_{p}/4)^{k}}%
{k!}e^{ik\pi\nu/\nu_{0}}. \label{Eq10}%
\end{equation}
Then, the Fourier transform of ES is
\begin{equation}
E_{0}(l,\nu)=E_{in}(\nu)e^{-d_{p}/4}\sum_{k=0}^{+\infty}\frac{(d_{p}/4)^{k}%
}{k!}e^{i\pi k\nu/\nu_{0}}, \label{Eq11}%
\end{equation}
The inverse Fourier transformation of $E_{0}(l,\nu)$%
\begin{equation}
E_{0}(l,t)=\frac{1}{2\pi}\int_{-\infty}^{+\infty}E_{0}(l,\nu)e^{-i\nu t}dt,
\label{Eq12}%
\end{equation}
gives the solution%
\begin{equation}
E_{\text{out}}(t)=e^{-d_{p}/4}\sum_{k=0}^{+\infty}\frac{(d_{p}/4)^{k}}%
{k!}E_{\text{in}}\left(  t-\frac{\pi k}{\nu_{0}}\right)  . \label{Eq13}%
\end{equation}
where $E_{\text{out}}(t)=E_{0}(l,t)$ is the field at the exit of HFC. Pulse
sequences, Eq. (\ref{Eq13}), generated at the exit of HFC with different
optical thickness of the absorption peaks $d_{p}$ are shown in Fig. 1. For
simplicity, as an input field a pulse with a Gaussian envelope is taken, which
is $E_{\text{in}}(t)=E_{0}\exp(-r^{2}t^{2})$. Parameter $r$ of the pulse is
equal $5\nu_{0}$, which means that the spectral width of the pulse covers more
than ten absorption peaks. With increase of the optical thickness $d_{p}$ of
the absorption peaks, the line, which links maximum amplitudes of the pulses,
generated at the exit of the crystal, forms a bell-shaped envelope, shown by
blue dotted line in Fig. 1(c). Its time dependence will be discussed in the
next section. \begin{figure}[ptb]
\resizebox{0.8\textwidth}{!}{\includegraphics{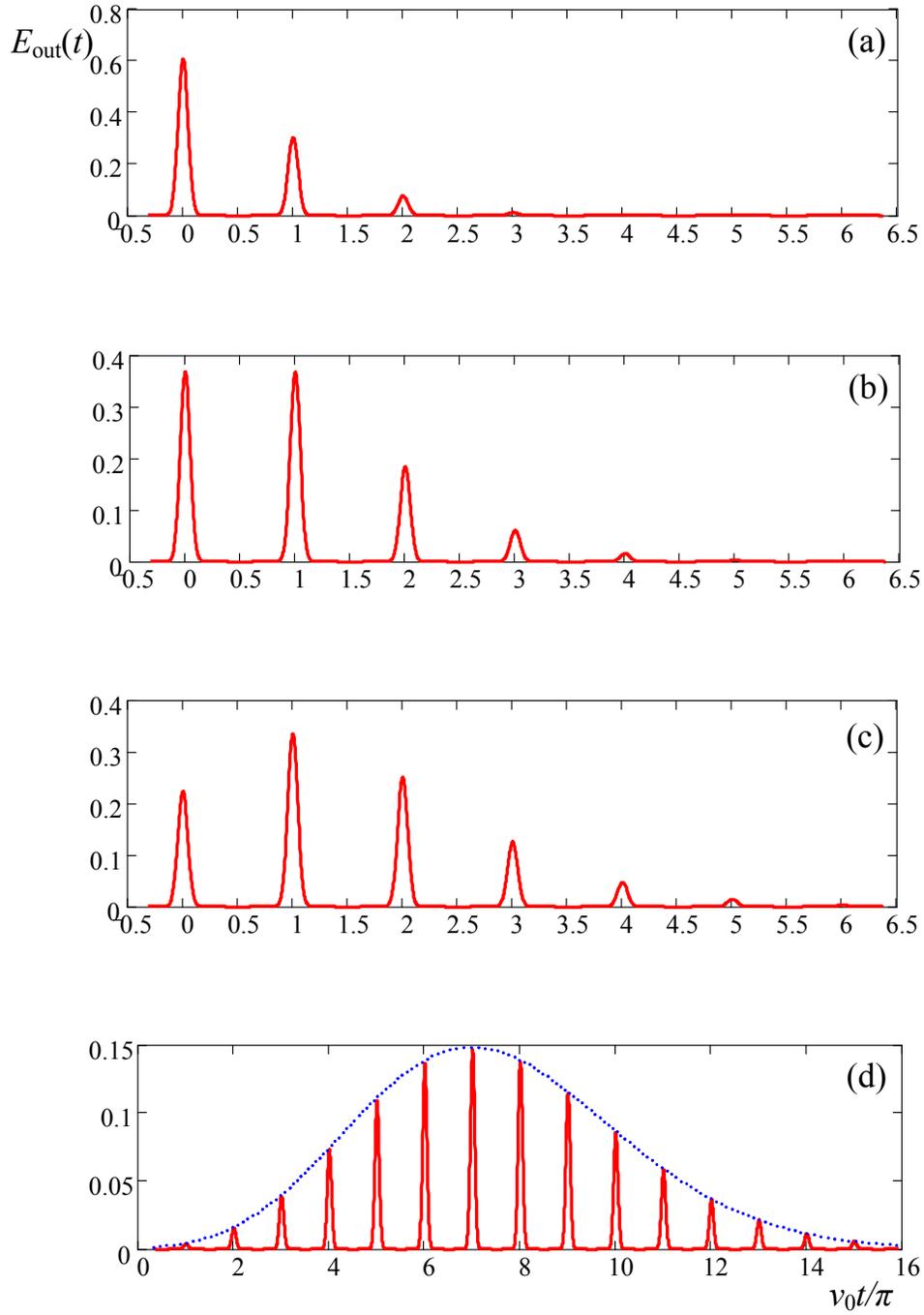}}\caption{Time
dependence of the output field form harmonic frequency crystal of different
optical thickness $d_{p}$, which is 2 (a), 4 (b), 6 (c), and 30 (d). Time
scale is in units $\pi/\nu_{0}$ and the field amplitude is normalized to the
maximum amplitude $E_{in}(0)$ of the input pulse. Input pulse has a Gaussian
shape $E_{in}(t)=E_{0}e^{-r^{2}t^{2}}$ with $r=5\nu_{0}$. Dotted line in blue
shows the net envelope of the pulse train, which is described by Eq.
(\ref{Eq25}), see Sec. IV. and discussion there.}%
\label{fig:1}%
\end{figure}

ES (\ref{Eq13}) is derived for the pulse whose central frequency $\omega_{c}$
is tuned in the center of one of the transparency windows. If there is a
frequency shift $\delta_{s}$ of $\omega_{c}$ with respect to the transparency
window center, the solution (\ref{Eq13}) is modified as
\begin{equation}
E_{\text{out}}(t)=e^{-d_{p}/4}\sum_{k=0}^{+\infty}\frac{(d_{p}/4)^{k}}%
{k!}E_{\text{in}}\left(  t-\frac{\pi k}{\nu_{0}}\right)  e^{i\pi k\delta
_{s}/\nu_{0}}. \label{Eq14}%
\end{equation}
When the central frequency of the pulse is tuned in resonance with one of the
absorption peaks, we have $\exp(i\pi k\delta_{s}/\nu_{0})=\exp(i\pi k)$ and
the pulses, generated at the exit of HFC at times $t_{k}=\pi k/\nu_{0}$ have
phase opposite to the phase of the input pulse if $k$ is odd, see Fig. 2 (a).
\begin{figure}[ptb]
\resizebox{0.6\textwidth}{!}{\includegraphics{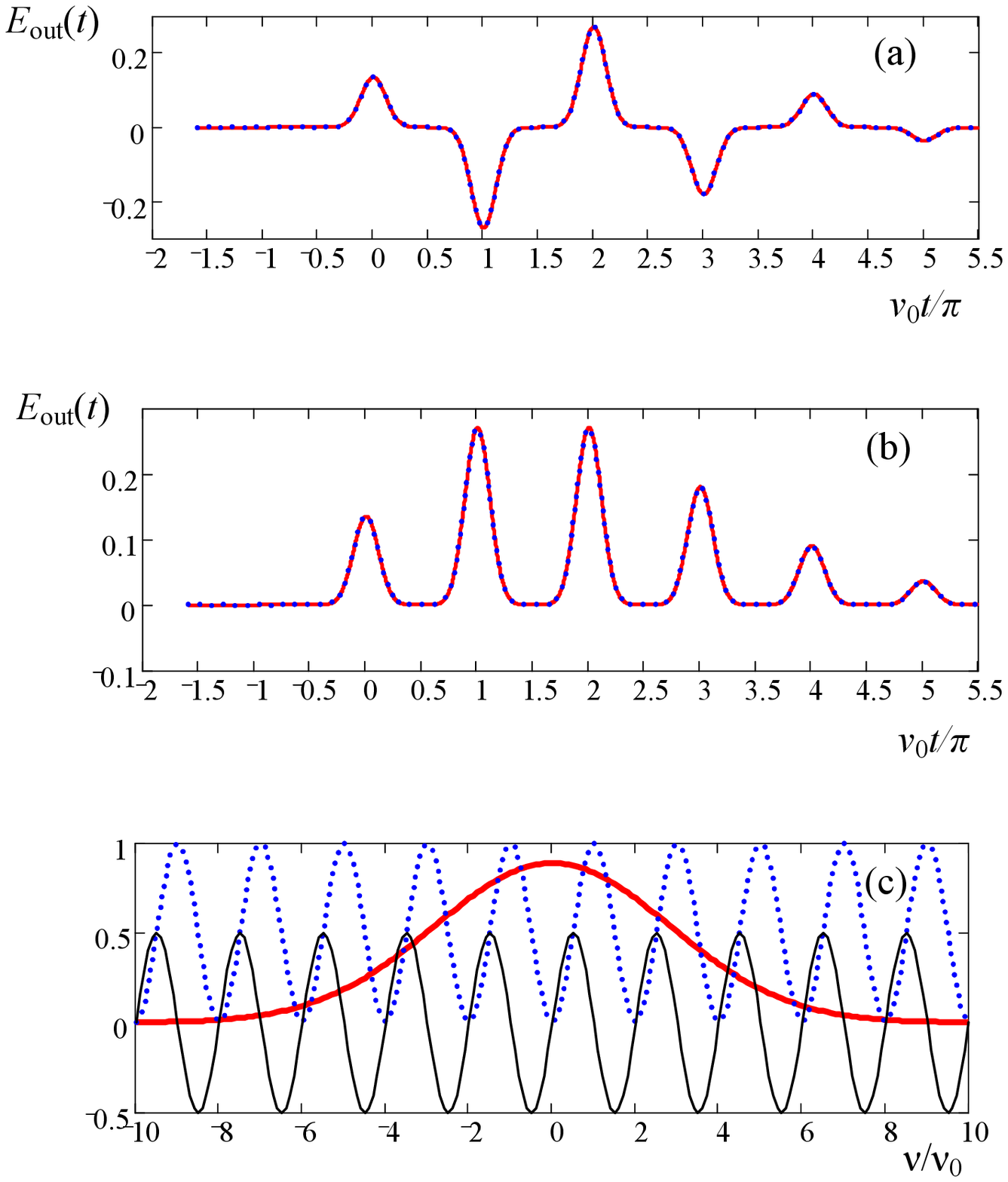}}\caption{Comparison of
the exact solution (solid line in red) with numerical solution (blue dotted
line), where the integration boundaries $\pm\Delta_{b}$ are limited by
$\pm10\nu_{0}$. Optical thickness of the absorption peaks is $d_{p}=8$. The
pulse central frequency is tuned in one of the absorption peaks (a) and in the
center of one of the transparency windows (b). Parameter $r$ of the input
pulse spectrum is equal to $2\nu_{0}$. Pulse amplitude is normalized to the
amplitude of the input pulse $E_{0}$. Absorption $\chi^{\prime\prime}(\nu)$
and dispersion $\chi^{\prime}(\nu)$ components of HFC are shown by dotted line
in blue and thin solid line in black in (c), respectively. Both are normalized
to $\chi_{0}$. Input pulse spectrum (multiplied by $\nu_{0}$ to make it
nondimentional) is shown by bold red line.}%
\label{fig:2}%
\end{figure}

Infinite spectrum of the frequency crystal is necessary only for the
derivation of the exact solution. Below we show that numerical solution, where
the spectrum of the pulse, transformed by the crystal, is integrated in the
limited domain, coincides with ES. The numerical solution is obtained by
calculating the integral in the equation%
\begin{equation}
E_{out}(t)=\frac{1}{2\pi}\int_{-\Delta_{b}}^{\Delta_{b}}E_{in}(\nu)e^{-i\nu
t-(d_{p}/4)[1\mp\cos\left(  \pi\nu/\nu_{0}\right)  \mp i\sin\left(  \pi\nu
/\nu_{0}\right)  ]}dt, \label{Eq15}%
\end{equation}
where $E_{in}(\nu)$ is the spectrum of the input pulse, $\Delta_{b}$ and
$-\Delta_{b}$ are the borders of the numerical integration, $\mp$ signs in the
exponent correspond to tuning the central frequency of the pulse in the center
of one of the transparency windows (sign minus) or in the center of one of the
absorption peaks (sign plus). Comparison of the exact solution with the
numerical results are shown in Fig. 2 (a,b) for the Gaussian pulse
$E_{\text{in}}(t)=E_{0}\exp(-r^{2}t^{2})$ whose spectrum is $E_{\text{in}}%
(\nu)=E_{0}(\sqrt{\pi}/r)\exp(-\nu^{2}/4r^{2})$. Parameter $r$ is taken equal
to $2\nu_{0}$. Integration boundaries ($\pm\Delta_{b}=\pm10\nu_{0}$) include
the contribution of $10$ absorption peaks, see Fig. 2 (c).

\section{Properties of the exact solution}

According to the exact solution (\ref{Eq13}) maximum intensity of the pulse,
transmitted through HFC with no delay $I_{\text{out}}(0)=\left\vert
E_{\text{out}}(0)\right\vert ^{2}$, is $I_{m0}=I_{m}\exp(-d_{p}/2)$ or
$I_{m0}=I_{m}\exp(-d_{p}/F)$, where $I_{m}=\left\vert E_{\text{in}%
}(0)\right\vert ^{2}$ is maximum intensity of the input pulse and $F=2$ is
finesse of HFC. Thus, maximum intensity of this pulse decreases according to
the Beers' law, $I_{m0}=I_{m}\exp(-d_{\text{eff}})$, with increase of the
effective optical thickness (optical depth) $d_{\text{eff}}=d_{p}/F$. This
thickness is reduced $F$ times with respect to the optical depth seen by the
monochromatic radiation, which is tuned in the center of the absorption peak.
This is because of transparency windows present in the spectrum. The result,
obtained for HFC, is consistent with properties of AFC with high finesse,
discussed in Refs.
\cite{Gisin2008,Gisin2009,Lauro,Bonarota,Chaneliere,Bonarota2012}.

Maximum intensity of the first pulse, delayed by time $t_{1}=\pi/\nu_{0}$, is
$I_{m1}=I_{m}(d_{\text{eff}}/2)^{2}\exp(-d_{\text{eff}})$, where
$I_{m1}=\left\vert E_{\text{out}}(t_{1})\right\vert ^{2}$. This relation also
supports the results, obtained in Refs.
\cite{Gisin2008,Gisin2009,Lauro,Bonarota,Chaneliere,Bonarota2012}. The
intensity $I_{m1}$\ takes maximum value $I_{m1}=0.135I_{m}$ when
$d_{\text{eff}}=2$. Maximum amplitude of the first pulse for this value of the
effective thickness is $E_{m1}(t_{1})=0.368E_{0}$, where $E_{0}=E_{\text{in}%
}(0)$, see Fig. 1 (b). Here and below, numbering of the pulse intensity
$I_{mk}$ and amplitude $E_{mk}(t_{k})$ follows the number $k$ in the pulse
delay $t_{k}=k\pi/\nu_{0}$.

The area of the output pulse train, shown in Fig. 1,
\begin{equation}
S_{\text{out}}=\int_{-\infty}^{+\infty}E_{\text{out}}(t)dt, \label{Eq16}%
\end{equation}
is conserved, $S_{\text{out}}=S_{\text{in}}$, (where $S_{\text{in}}$ is the
input pulse area) if the central frequency of the input pulse is tuned in the
center of one of the transparency windows. For the input pulse, whose central
frequency is tuned in resonance with one of the absorption peaks, pulse-train
area reduces as $S_{\text{out}}=S_{\text{in}}\exp(-d_{\text{eff}})$. These
results follow from ES (\ref{Eq13}) and (\ref{Eq14}) since
\begin{equation}
\sum_{k=0}^{+\infty}\frac{(\pm d_{\text{eff}}/2)^{k}}{k!}=e^{\pm
d_{\text{eff}}/2}. \label{Eq17}%
\end{equation}
Time integrated intensity of the output pulses,
\begin{equation}
\left\langle I_{\text{out}}\right\rangle _{t}=\int_{-\infty}^{+\infty
}\left\vert E_{\text{out}}(t)\right\vert ^{2}dt, \label{Eq18}%
\end{equation}
decreases as%
\begin{equation}
\left\langle I_{\text{out}}\right\rangle _{t}=e^{-d_{\text{eff}}}%
I_{0}(d_{\text{eff}})\left\langle I_{\text{in}}\right\rangle _{t},
\label{Eq19}%
\end{equation}
irrespective to the pulse central frequency $\omega_{c}$, since%
\begin{equation}
\sum_{k=0}^{+\infty}\frac{(\pm d_{\text{eff}}/2)^{2k}}{(k!)^{2}}%
=I_{0}(d_{\text{eff}}), \label{Eq20}%
\end{equation}
where $\left\langle I_{\text{in}}\right\rangle _{t}$ is the time integrated
intensity of the input pulse and $I_{0}(x)$ is the modified Bessel function of
zero order, see Ref. \cite{Abramowitz}. When $d_{\text{eff}}\geqslant3.75$,
Eq. (\ref{Eq19}) is approximated as
\begin{equation}
\left\langle I_{\text{out}}\right\rangle _{t}=\frac{0.4}{\sqrt{d_{\text{eff}}%
}}\left\langle I_{\text{in}}\right\rangle _{t}. \label{Eq21}%
\end{equation}
For $0\leqslant d_{\text{eff}}\leqslant3.75$, Eq. (\ref{Eq19}) is approximated
by expression%
\begin{equation}
\left\langle I_{\text{out}}\right\rangle _{t}=(1+0.25d_{\text{eff}}%
^{2}+0.016d_{\text{eff}}^{4}+0.0004d_{\text{eff}}^{6})e^{-d_{\text{eff}}%
}\left\langle I_{\text{in}}\right\rangle _{t}, \label{Eq22}%
\end{equation}
see Ref. \cite{Abramowitz}. Graphical illustration of the dependence
(\ref{Eq22}) on $d_{\text{eff}}$ is shown in Fig.3. \begin{figure}[ptb]
\resizebox{0.5\textwidth}{!}{\includegraphics{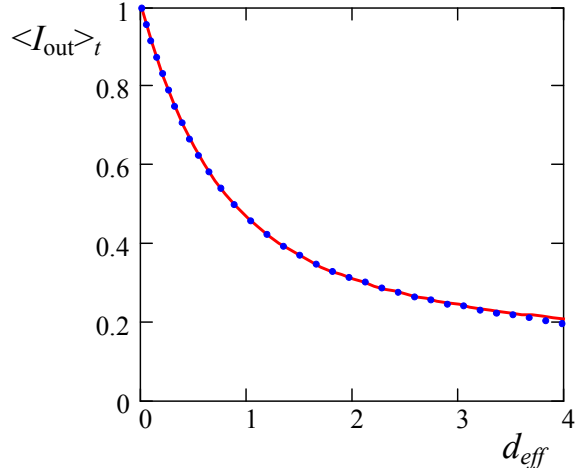}}\caption{Dependence of
the time integrated intensities of the pulses (net intensity) at the exit of
HFC (normalized to $\left\langle I_{\text{in}}\right\rangle _{t}$) on
effective thickness $d_{\text{eff}}$. Solid line in read is the exact
dependence, described by Eq. (\ref{Eq19}). Dotted line in blue is the
approximation, given by Eq. (\ref{Eq22}).}%
\label{fig:3}%
\end{figure}

If effective thickness is large ($d_{\text{eff}}\gg1$), the maximum amplitudes
of the pulses form a bell-shaped envelope [see Fig. 1 (d)]. The envelope can
be described analytically if we use Stirling's formula for factorial, which is%
\begin{equation}
k!=\sqrt{2\pi}k^{k+1/2}e^{-k+\theta/12k}, \label{Eq23}%
\end{equation}
where $k>0$ and $0<\theta<1$. This formula helps to derive the approximate
dependence of the maximum amplitude of the $k$th pulse, $E_{mk}(t_{k})$, on
$k$ as
\begin{equation}
E_{mk}(t_{k})=e^{-d_{p}/4}\frac{(d_{p}/4)^{k}}{k!}E_{0}\approx e^{-d_{p}%
/4}\frac{(ed_{p}/4k)^{k}}{\sqrt{2\pi k}}E_{0}, \label{Eq24}%
\end{equation}
see Eq. (\ref{Eq13}). Taking into account that $k$th pulse reaches its maximum
amplitude at time $t_{k}=\pi k/\nu_{0}$, we express number $k$ through time
$t_{k}$, i.e., $k=\nu_{0}t_{k}/\pi$, and obtain
\begin{equation}
E_{mk}(t_{k})\approx e^{-d_{p}/4}\frac{(\pi ed_{p}/4\nu_{0}t_{k})^{\nu
_{0}t_{k}/\pi}}{\sqrt{2\nu_{0}t_{k}}}E_{0}. \label{Eq25}%
\end{equation}
Then, we allow time $t_{k}$ to evolve continuously in Eq. (\ref{Eq25}) and
obtain the function $E_{m}(t)$, where index $k$ is omitted. The plot of this
function is shown in Fig. 1 (d) by dotted line in blue. For large $d_{p}$, the
function $E_{m}(t)$ describes excellent the net envelope of the pulse train.
The maximum of this envelope takes place at time $t_{\max}=\pi(d_{p}%
-2)/4\nu_{0}$, which is found from the condition $\partial E_{m}(t)/\partial
t=0$. In the example, shown in Fig. 1 (d) for $d_{p}=30$, the envelope maximum
is formed when $\nu_{0}t_{\max}/\pi=7$.

For large $d_{\text{eff}}$ the maximum of the envelope, $E_{\max}$, at time
$t_{\max}$, is described by equation
\begin{equation}
E_{\max}=\sqrt{\frac{1}{\pi d_{\text{eff}}F(d_{\text{eff}})}}, \label{Eq26}%
\end{equation}
where%
\begin{equation}
F(d_{\text{eff}})=e\left(  1-\frac{1}{d_{\text{eff}}}\right)  ^{d_{\text{eff}%
}}. \label{Eq27}%
\end{equation}
According to the notable special limit%
\begin{equation}
\lim_{d_{\text{eff}}\rightarrow+\infty}\left(  1-\frac{1}{d_{\text{eff}}%
}\right)  ^{d_{\text{eff}}}=e^{-1}, \label{Eq28}%
\end{equation}
the function $F(d_{\text{eff}})$ tends to $1$. This function differs from $1$
very little if $d_{\text{eff}}\geq10$ (i.e., $F(10)=0.891$) and takes values
not very differen from $1$ for smaller $d_{\text{eff}}$. For example, we have
$F(5)=0.758$.

Halfwidth at halfmaximum of the function $E_{m}(t)$ is found from the equation
$E_{m}(t_{\max}\pm t_{\text{h}\max})=E_{m}(t_{\max})/2$. Its solution gives
full width at half maximum $t_{\text{w}}=2t_{\text{h}\max}$, which is
approximated as
\begin{equation}
t_{\text{w}}=\frac{\pi}{\nu_{0}}\sqrt{d_{\text{eff}}2\ln(2)} \label{Eq29}%
\end{equation}
for large value of the optical thickness $d_{\text{eff}}$.

\section{Slow light}

Exact solution (\ref{Eq13}) is universal and capable to describe pulse
propagation as a whole with reduced group velocity. In this section we
consider the propagation of the Gaussian pulse $E_{\text{in}}(t)=E_{0}%
\exp(-r^{2}t^{2})$ whose central frequency $\omega_{c}$ is tuned in the center
of one of the transparency windows. The spectrum of the pulse is
$E_{\text{in}}(\nu)=E_{0}(\sqrt{\pi}/r)\exp(-\nu^{2}/4r^{2})$. We consider the
case when the spectral width of the pulse is much smaller than the width of
the transparency window, i.e., $r\ll2\nu_{0}$. ES solution for this pulse is
shown in Fig. 4.

Following approach, developed in Refs. \cite{Shakhmuratov1,Shakhmuratov2}, one
can use expansion in a power series of the complex dielectric constant in the
exponent of Eq. (\ref{Eq9}), which gives the following expression for the
transmission function
\begin{equation}
T_{\mathrm{h}}(\nu)=\frac{d_{\text{eff}}}{2}\left(  1-e^{i\pi\nu/\nu_{0}%
}\right)  \approx\frac{d_{\text{eff}}}{2}\left[  -i\frac{\pi\nu}{\nu_{0}%
}+\frac{1}{2}\left(  \frac{\pi\nu}{\nu_{0}}\right)  ^{2}+...\right]  .
\label{Eq30}%
\end{equation}
The first term in the square brackets gives delay of the pulse due to the
reduced group velocity. The second term describes pulse broadening in time and
its spectrum narrowing with propagation distance due to the absorbtion of the
wings of the pulse spectrum \cite{Shakhmuratov1,Shakhmuratov2}. Analytical
calculation of the inverse Fourier transformation (\ref{Eq12}) for
$E_{0}(l,\nu)$ with the approximated transmission function $T_{\mathrm{h}}%
(\nu)$, Eq. (\ref{Eq30}), gives
\begin{equation}
E_{\text{out}}(t)=\frac{E_{0}}{w_{a}}\exp\left[  -r_{m}^{2}\left(
t-t_{D}\right)  ^{2}\right]  ), \label{Eq31}%
\end{equation}
where $t_{D}=$ $d_{\text{eff}}\pi/2\nu_{0}$ is a delay time due to the reduced
group velocity, $r_{m}=r/w_{a}$ is a parameter which describes $w_{a}$ times
broadening of the pulse in time and%
\begin{equation}
w_{a}=\sqrt{1+d_{\text{eff}}\left(  \frac{\pi r}{\nu_{0}}\right)  ^{2}}.
\label{Eq32}%
\end{equation}
Comparison of the approximate solution (\ref{Eq31}) with the exact solution is
shown in Fig. 4 for $r=\nu_{0}/20$. The difference between them is almost
negligible. \begin{figure}[ptb]
\resizebox{0.6\textwidth}{!}{\includegraphics{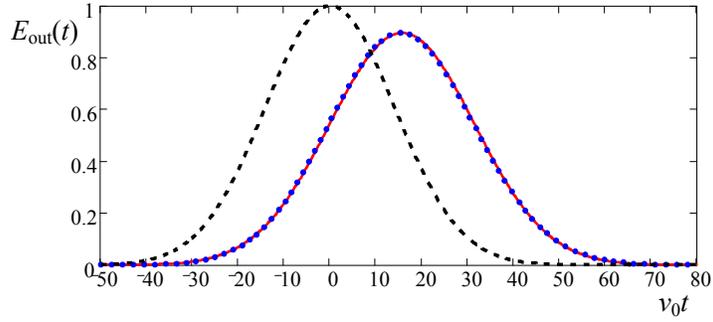}}\caption{Time
dependence of the Gaussian pulse at the input of the medium (black dashed
line) and at the exit of the frequency crystal with effective optical
thickness $d_{\text{eff}}=10$. Solid line in read is the exact solution and
blue dotted line is the approximate analytical solution, Eq. (\ref{Eq31}).
Time scale is given in units $\nu_{0}$. Parameter $r$ of the Gaussian pule is
$r=\nu_{0}/20$. Pulse amplitude is normalized to $E_{0}$.}%
\label{fig:4}%
\end{figure}

\section{Resonant absorption}

If the central frequency of the pulse is tuned in resonance with one of the
absorption peaks and its spectral width is smaller than the width of the
absorption peak, the pulse propagates in the frequency crystal as in the
absorptive medium. Exact solution for the amplitude of the pulse in resonance
with the absorption peaks irrespective to the spectral width of the pulse is
described by
\begin{equation}
E_{\text{out}}(t)=e^{-d_{p}/4}\sum_{k=0}^{+\infty}(-1)^{k}\frac{(d_{p}/4)^{k}%
}{k!}E_{\text{in}}\left(  t-\frac{\pi k}{\nu_{0}}\right)  . \label{Eq33}%
\end{equation}
Time dependence of the pulse at the exit of the frequency crystals of
different optical thickness is shown in Fig. 5. Qualitatively, attenuation of
the pulse and transformation of its shape are very similar to those, which are
typical for the pulse propagating in a thick two-level medium (cf. with Figs.
3 and 4 in Ref. \cite{Crisp}). \begin{figure}[ptb]
\resizebox{0.7\textwidth}{!}{\includegraphics{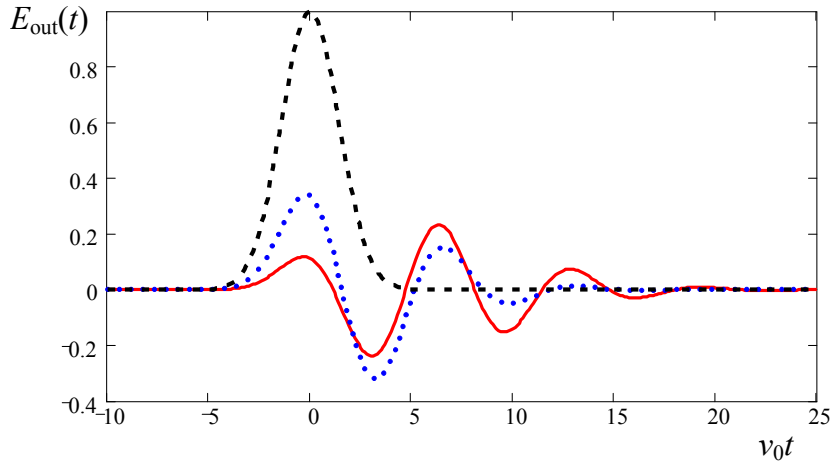}}\caption{Time
dependence of the Gaussian pulse at the input of the medium (black dashed
line) and at the exit of the frequency crystal with effective optical
thickness $d_{\text{eff}}$ equal 2 (dotted line in blue) and 4 (solid line in
red). Time scale is in units $\nu_{0}$. Parameter $r$ of the Gaussian pule is
$r=\nu_{0}/2$. Pulse amplitude is normalized to $E_{0}$.}%
\label{fig:5}%
\end{figure}

\section{Anharmonic frequency crystal}

Another example of the atomic frequency comb is an anharmonic frequency
crystal (AHFC). It has also a periodic structure in the absorption spectrum
with a period $2\nu_{0}$ but many multiple frequencies $k\nu_{0}$ (where $k$
is integer) contribute to the spectrum. Below we consider the crystal with the
imaginary part of the susceptibility, which is described by the function
\begin{equation}
\chi_{\mathrm{ah}}^{\prime\prime}(\nu)=\chi_{0}\sin^{10}\left(  \frac{\pi\nu
}{2\nu_{0}}\right)  .\label{Eq34}%
\end{equation}
Its counterpart, the real part of the susceptibly, is found from the
Kramers-Kronig relation, Eq. (\ref{Eq4}), which gives
\begin{equation}
\chi_{\mathrm{ah}}^{\prime}(\nu)=\frac{\chi_{0}}{2^{9}}\sum_{k=0}^{4}%
(-1)^{k}C_{10}^{k}\sin\left[  \frac{(5-k)\pi\nu}{\nu_{0}}\right]
.\label{Eq35}%
\end{equation}
where $C_{10}^{k}$ is the binomial coefficient,
\begin{equation}
C_{10}^{k}=\frac{10!}{(10-k)!k!}.\label{Eq36}%
\end{equation}
This result follows from the binomial theorem
\begin{equation}
\sin^{10}\left(  \frac{\pi\nu}{2\nu_{0}}\right)  =\left(  i\frac{e^{-i\pi
\nu/2\nu_{0}}-e^{i\pi\nu/2\nu_{0}}}{2}\right)  ^{10}=-2^{-10}\sum_{k=0}%
^{10}(-1)^{k}C_{10}^{k}e^{i(5-k)\pi\nu/\nu_{0}}.\label{Eq37}%
\end{equation}
Frequency dependencies of the absorption $\sim\chi^{\prime\prime}(\nu)$ and
dispersion $\sim\chi^{\prime}(\nu)$\ of the anharmonic frequency crystal are
shown in Fig. 6. They have a periodic dependence on frequency $\nu$ with a
period $2\nu_{0}$. However, according to Eqs. (\ref{Eq35}) and (\ref{Eq37}),
except the oscillations with frequency $\nu_{0}$, they have also contribution
of the harmonics $2\nu_{0}$, $3\nu_{0}$, $4\nu_{0}$, and $5\nu_{0}$. The width
at half maximum of the absorption peaks is $4.27$ times smaller that the
distance between them, which gives finesse $F=4.27$. \begin{figure}[ptb]
\resizebox{0.5\textwidth}{!}{\includegraphics{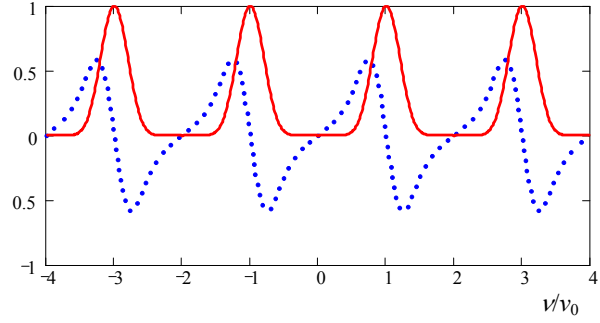}}\caption{Absorption,
$\chi^{\prime\prime}(\nu)$, (red solid line) and dispersion, $\chi^{\prime
}(\nu)$, (blue dotted line) components of the anharmonic frequency crystal,
described by Eqs. (\ref{Eq34}) and (\ref{Eq35}). Frequency scale is in units
$\nu_{0}$.}%
\label{fig:6}%
\end{figure}The complex dielectric constant in the exponent of Eq. (\ref{Eq8})
gives the following expression for the transmission function of the AHFC
\begin{equation}
T_{\mathrm{ah}}(\nu)=\frac{d_{p}}{2\chi_{0}}\left[  \chi_{\mathrm{ah}}%
^{\prime\prime}(\nu)-i\chi_{\mathrm{ah}}^{\prime}(\nu)\right]  ,\label{Eq38}%
\end{equation}
which can be expressed as%
\begin{equation}
T_{\mathrm{ah}}(\nu)=\frac{d_{p}}{2^{10}}\left[  \sum_{k=1}^{5}(-1)^{k+1}%
C_{10}^{5-k}e^{ik\pi\nu/\nu_{0}}+\frac{C_{10}^{5}}{2}\right]  ,\label{Eq39}%
\end{equation}
where $d_{p}$ is the optical thickness of the crystal for the monochromatic
radiation field tuned in resonance with one of the absorption peaks.

With the help of the expansion of the exponent in the equation
\begin{equation}
E_{0}(l,\nu)=E_{0}(0,\nu)\exp\left[  -T_{\mathrm{ah}}(\nu)\right]  ,
\label{Eq40}%
\end{equation}
in a power series of $T_{\mathrm{ah}}(\nu)$ and inverse Fourier
transformation, Eq. (\ref{Eq12}), one can derive analytical time dependence of
at least first three pulses at the exit of the anharmonic frequency crystal,
i.e.,%
\begin{equation}
E_{\text{out}}(t)=e^{-63d_{p}/512}\left[  E_{\text{in}}\left(  t\right)
+A_{1}(d_{p})E_{\text{in}}\left(  t-\frac{\pi}{\nu_{0}}\right)  +A_{2}%
(d_{p})E_{\text{in}}\left(  t-\frac{2\pi}{\nu_{0}}\right)  +...\right]  ,
\label{Eq41}%
\end{equation}
where%
\begin{equation}
A_{1}(d_{p})=\frac{105}{512}d_{p}, \label{Eq42}%
\end{equation}
and%
\begin{equation}
A_{2}(d_{p})=-\frac{15}{128}d_{p}+\frac{1}{2}\left(  \frac{105}{512}\right)
^{2}d_{p}^{2}. \label{43}%
\end{equation}
\begin{figure}[ptb]
\resizebox{0.7\textwidth}{!}{\includegraphics{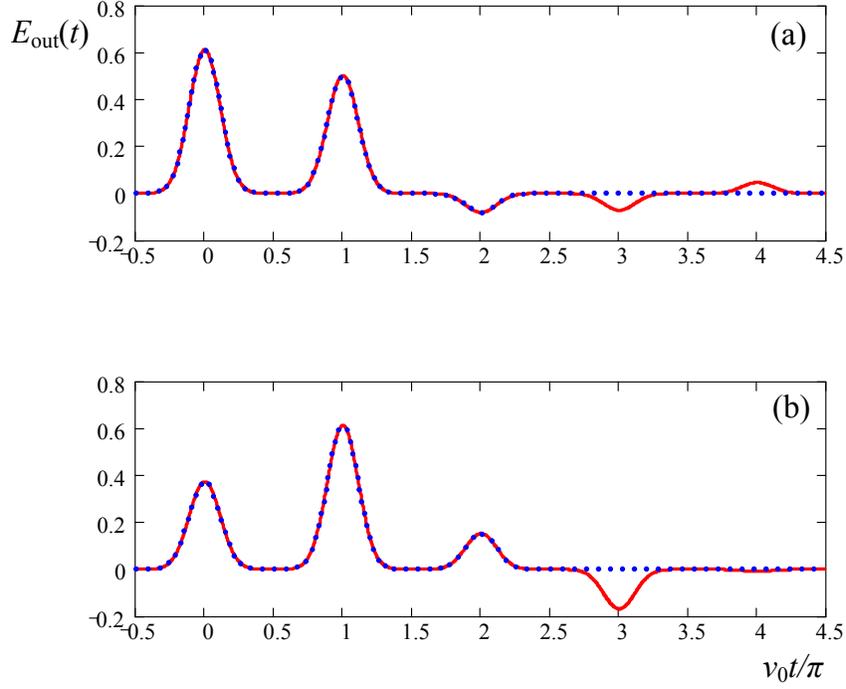}}\caption{Time
dependence of the Gaussian pulse transmitted through AHFC with optical
thickness $d_{p}=4$ (a) and 8 (b). Parameter $r$ of the Gaussian pulse is
$2\nu_{0}$. Solid line in red corresponds to the numerical integration
according to Eq. (\ref{Eq12}) with the transmission function, defined in Eqs.
(\ref{Eq39}) and Eq. (\ref{Eq40}). Dotted line in blue is plotted according to
the analytical result, given in Eq. (\ref{Eq41}), where only first three terms
are taken into account.}%
\label{fig:7}%
\end{figure}\begin{figure}[ptbptb]
\resizebox{0.4\textwidth}{!}{\includegraphics{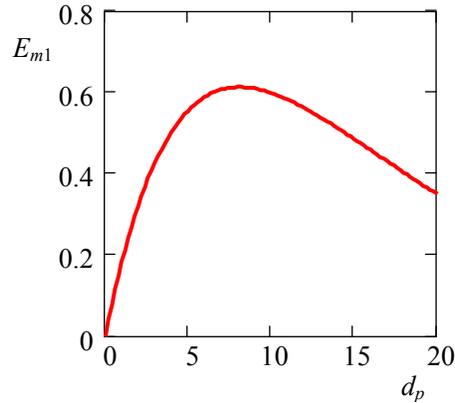}}\caption{Dependence of
the maximum amplitude of the first delayed pulse on the optical thickness
$d_{p}$.}%
\label{fig:8}%
\end{figure}Comparison of the time dependencies of the numerically calculated
$E_{\text{out}}(t)$ with the help of Eqs. (\ref{Eq12}),(\ref{Eq40}) and
analytical approximation, Eq. (\ref{Eq41}), where only first three pulses are
taken into account, is shown in Fig. 7. Coincidence for the first three pulses
is excellent.

Exact expression for the maximum amplitude of the pulse with no delay is
$E_{0}\exp(-63d_{p}/512)$, which gives $E_{0}\exp(-d_{p}/2F_{\mathrm{exc}})$,
where $F_{\mathrm{exc}}=4.063$. This value is slightly smaller than finesse
$F=4.27$, estimated from the structure of the absorption spectrum.

Exact expression for the maximum amplitude of the first delayed pulse is
$E_{0}A_{1}(d_{p})\exp(-63d_{p}/512)$. Its dependence on the optical thickness
$d_{p}$ is shown in Fig. 8. Maximum amplitude of this pulse, $0.613E_{0}$, is
achieved when $63d_{p}/512=1$, which gives $d_{p}\simeq8$. For this value of
thickness we have $E_{m1}=E_{0}105/(63e)$.

\section{Hole burning technique}

Frequency filters with a periodic spectrum could be realized, for example,
with the help of the combination of the system of many mini-cavities
interacting with a common broadband cavity coupled with the external waveguide
\cite{Moiseev}. However, most common technique is the hole burning, which
allows to create periodic sequence of the absorption lines and transparent
windows in a wide inhomogeneously broadened absorption spectrum
\cite{Gisin2008,Gisin2009,Lauro,Bonarota,Chaneliere,Bonarota2012}. In the
preparation stage the periodic dependence of the population difference
$n(\omega_{A})$ of the ground and excited states on resonance frequency of
atoms $\omega_{A}$ is created. Below, the real and imaginary parts of the
atomic susceptibility are derived for HFC and two types of AHFC.

For a weak pulse the linear response approximation (LRA) is applicable for the
description of the density matrix evolution of atoms in the medium, $\rho
_{mn}(z,t)$. The population change of the ground $g$ and excited $e$ states is
neglected in LRA and only the equation for the nondiagonal element, $\rho
_{eg}(z,t)=\sigma_{eg}(z,t)\exp(-i\omega_{c}t+ik_{c}z)$, is considered in the
form
\begin{equation}
\frac{\partial}{\partial t}\sigma_{eg}(z,t)=(i\Delta-\gamma)\sigma
_{eg}(z,t)+i\Omega(z,t)n(\Delta), \label{Eq44}%
\end{equation}
where $\omega_{c}$ and $k_{c}$ are the frequency and the wave number of the
input pulse, $z$ is the propagation distance, counted from the input face of
the medium inside, $\gamma$ is the decay rate of the atomic coherence
responsible for the homogeneous broadening of the absorption line,
$\sigma_{eg}(t)$ is the slowly varying part of the nondiagonal element of the
atomic density matrix, $\Delta=\omega_{c}-\omega_{A}$ is the difference of the
frequency $\omega_{c}$ of the weak pulsed field and resonant frequency
$\omega_{A}$ of an individual atom, $\Omega(t)=d_{eg}E_{0}(z,t)/2\hbar$ is the
Rabi frequency, which is proportional to the time varying field amplitude
$E_{0}(z,t)$ and dipole-transition matrix element between $g$ and $e$ states,
$d_{eg}$, and $n(\Delta)$ is the long-lived population difference, created by
the hole burning. If atom is in the ground state, then $n(\Delta)$ is equal
unity. If atom with the frequency $\omega_{A}=\omega_{c}-\Delta$ is removed by
the hole burning to the shelving state, then $n(\Delta)$ is zero.

With the help of the Fourier transform, Eq. (\ref{Eq44}) is reduced to the
algebraic equation whose solution is%
\begin{equation}
\sigma_{eg}(z,\nu)=-\frac{\Omega(z,\nu)n(\Delta)}{\nu+\Delta+i\gamma}.
\label{Eq45}%
\end{equation}
In the slowly varying amplitude approximation the wave equation is reduced to%
\begin{equation}
\widehat{L}E_{0}(z,t)=i\hbar\alpha\gamma\sigma_{eg}(z,t)/d_{eg}, \label{Eq46}%
\end{equation}
where $L=\partial_{z}+c^{-1}\partial_{t}$, $\alpha_{p}=4\pi\omega
_{c}N\left\vert d_{eg}\right\vert ^{2}/\gamma\hbar c$ is the absorption
coefficient, and $N$ is the density of atoms. After Fourier transformation the
wave equation is reduced to%
\begin{equation}
\left[  \frac{\partial}{\partial z}-\frac{i\nu}{c}+\alpha(\nu)\right]
\Omega(z,\nu)=0, \label{Eq47}%
\end{equation}
where%
\begin{equation}
\alpha(\nu)=\frac{i\alpha_{p}\gamma/2}{\nu+\Delta+i\gamma}. \label{Eq48}%
\end{equation}
Solution of Eq. (\ref{Eq47}),%
\begin{equation}
E_{0}(z,\nu)=E_{0}(0,\nu)\exp[(i\nu z/c)-\alpha(\nu)z], \label{Eq49}%
\end{equation}
coincides with Eq. (\ref{Eq8}).

Below, HFC and two examples of AHFC will be considered.

\subsection{Harmonic Frequency Crystal}

We consider the case when population difference follows harmonic frequency
dependence, i.e.,%
\begin{equation}
n(\Delta)=\frac{1-\cos(\pi\Delta/\nu_{0})}{2}. \label{Eq50}%
\end{equation}
For simplicity, we suppose that before the hole burning the frequency
distribution of atoms was infinite and uniform. This assumption corresponds to
a broad and flat inhomogeneous broadening, which is valid if inhomogeneous
width is much larger than the period $2\nu_{0}$. Then the averaged atomic
coherence is%
\begin{equation}
\left\langle \sigma_{eg}(\nu)\right\rangle _{\Delta}=\frac{1}{\pi\Gamma}%
{\displaystyle\int\nolimits_{-\infty}^{+\infty}}
\sigma_{eg}(\nu)d\Delta, \label{Eq51}%
\end{equation}
where $\Gamma$ is the width of inhomogeneous broadening. This expression
cannot be used to estimate exact value of the optical thickness $d_{p}$ since
it does not contain the information about the width and shape of the initial
inhomogeneous broadening and the way how the periodic structure in the
absorption spectrum was created. However, it allows to find the actual
frequency dependencies of $\chi^{\prime}(\nu)$, $\chi^{\prime\prime}(\nu)$ and
derive the solutions (\ref{Eq9}) and (\ref{Eq11}) for the case of the hole burning.

Analytical calculation of the integral (\ref{Eq51}) within adopted
approximation gives%
\begin{equation}
\left\langle \sigma_{eg}(\nu)\right\rangle _{\Delta}=i\frac{\Omega(\nu
)}{2\Gamma}\left(  1-e^{-\pi\gamma/\nu_{0}+i\pi\nu/\nu_{0}}\right)  .
\label{Eq52}%
\end{equation}
From this result it follows that if population difference has harmonic
frequency dependence, Eq. (\ref{Eq50}), then the Fourier transform of the
field at the exit of the medium is described by
\begin{equation}
E_{0}(l,\nu)=E_{0}(0,\nu)\exp\left\{  i\nu\frac{l}{c}-\frac{d_{p}}{4}%
+\frac{d_{r}}{4}e^{-\pi\gamma/\nu_{0}+i\pi\nu/\nu_{0}}\right\}  , \label{Eq53}%
\end{equation}
cf. Eq. (\ref{Eq9}). Thus, harmonic frequency dependence of the population
difference results in a decrease of the parameter $d_{p}$ to $d_{r}=d_{p}%
\exp(-\pi\gamma/\nu_{0})$ in the expression for all the time delayed pulses
and does not affect the amplitude of the pulse with no delay, i.e.,%
\begin{equation}
E_{0}(l,\nu)=E_{in}(\nu)e^{-d_{p}/4}\sum_{k=0}^{+\infty}\frac{(d_{r}/4)^{k}%
}{k!}e^{i\pi k\nu/\nu_{0}}, \label{Eq54}%
\end{equation}
If $\gamma\ll\nu_{0}$, this decrease is negligible.

\subsection{Anharmonic Frequency Crystal I}

For the anharmonic frequency crystal, considered in Sec. VII, the population
difference has a frequency dependence%
\begin{equation}
n(\Delta)=\sin^{10}(\pi\Delta/\nu_{0}). \label{Eq55}%
\end{equation}
After averaging in accord with Eq. (\ref{Eq51}), one obtains%
\begin{equation}
\left\langle \sigma_{eg}(\nu)\right\rangle _{\Delta}=i\frac{\Omega(\nu
)}{2^{11}\Gamma}\left[  \frac{C_{10}^{5}}{2}+\sum_{k=1}^{5}(-1)^{k+1}%
C_{10}^{5-k}e^{k\pi(i\nu-\gamma)/\nu_{0}}\right]  . \label{Eq56}%
\end{equation}
Similar to the harmonic frequency crystal, the produced pulses are described
by almost the same equation as in Sec. VII, but with the reduced amplitudes of
the delayed pulses, i.e.,%
\begin{equation}
A_{1}(d_{p})=\frac{105}{512}d_{r}, \label{Eq57}%
\end{equation}%
\begin{equation}
A_{2}(d_{p})=-\frac{15}{128}d_{r}e^{-\pi\gamma/\nu_{0}}+\frac{1}{2}\left(
\frac{105}{512}\right)  ^{2}d_{r}^{2}, \label{Eq58}%
\end{equation}
where $d_{r}=d_{p}\exp(-\pi\gamma/\nu_{0})$.

\subsection{Anharmonic Frequency Crystal II}

In this section a periodic structure of Lorentzian peaks is considered. It can
be classified as the anharmonic frequency crystal or AHFC II. This structure
is possible to construct by repumping a set of Lorentzian peaks into a broad
absorption dip, created initially by the hole burning, see Ref.
\cite{Chaneliere} where such an AFC was considered. For the numerical analysis
we consider the finite sequence of the Lorentzian peaks in the population
difference, which is descibed by%
\begin{equation}
n(\Delta)=\sum_{n=-101}^{100}\frac{\Gamma_{L}^{2}}{[\Delta+\nu_{L}%
(2n+1)]^{2}+\Gamma_{L}^{2}}, \label{Eq59}%
\end{equation}
where $\Gamma_{L}$ is a halfwidth of the Lorentzian peak and $2\nu_{L}$ is a
distance between two neighboring peaks. The limits in the sum are chosen such
that the absorption peaks are symmetrically placed around the central
transparency window.

Calculating the integral in Eq. (\ref{Eq51}), one obtains%
\begin{equation}
\left\langle \sigma_{eg}(\nu)\right\rangle _{\Delta}=-\frac{\Omega(\nu
)}{2\Gamma}\sum_{n=-101}^{100}\frac{\Gamma_{L}}{\nu-\nu_{L}(2n+1)+i(\Gamma
_{L}+\gamma)}. \label{Eq60}%
\end{equation}
This result allows to express the Fourier transform of the pulses at the exit
of the medium as%
\begin{equation}
E_{0}(l,\nu)=E_{0}(0,\nu)\exp\left[  -T_{L}(\nu)\right]  , \label{Eq61}%
\end{equation}
where%
\begin{equation}
T_{L}(\nu)=\frac{d_{p}}{2}\sum_{n=-101}^{100}\frac{i\Gamma_{0}}{\nu-\nu
_{L}(2n+1)+i\Gamma_{0}}, \label{Eq62}%
\end{equation}
and $\Gamma_{0}=\Gamma_{L}+\gamma$.

To compare AHFC consisting of the periodic sequence of the Lorentzian
absorption peaks (AHFC II) with that considered in the previous subsection
(AHFC I), we select such a value of $\Gamma_{0}$ when absorption spectra of
both are very similar. This condition is more or less satisfied when
$\Gamma_{0}=0.157\nu_{L}$ and the period of both structures coincide, i.e.,
$\nu_{0}=\nu_{L}$, see Fig. 9.\begin{figure}[ptb]
\resizebox{0.6\textwidth}{!}{\includegraphics{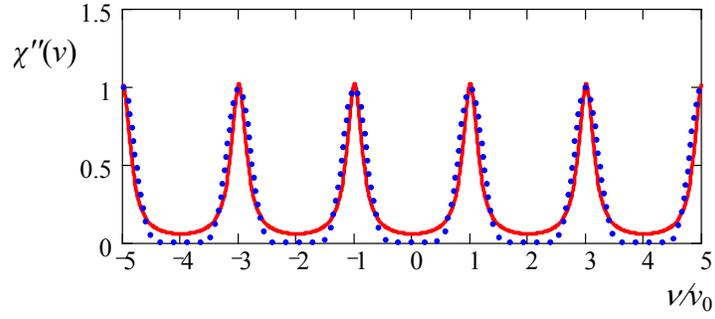}}\caption{Comarison of
the absorption peaks, which are described by the imaginary part of the
susceptibility, $\chi^{\prime\prime}(\nu)$, for the AHFC II consisting of the
Lorentzian peaks (red solid line) and AHFC I described by the function
$\sin^{10}(\pi\Delta/\nu_{0})$ (dotted line in blue). The frequency
dependencies are plotted for $\nu_{L}=\nu_{0}$ and $\Gamma_{L}=0.157\nu_{0}$.}%
\label{fig:9}%
\end{figure}\begin{figure}[ptbptb]
\resizebox{0.8\textwidth}{!}{\includegraphics{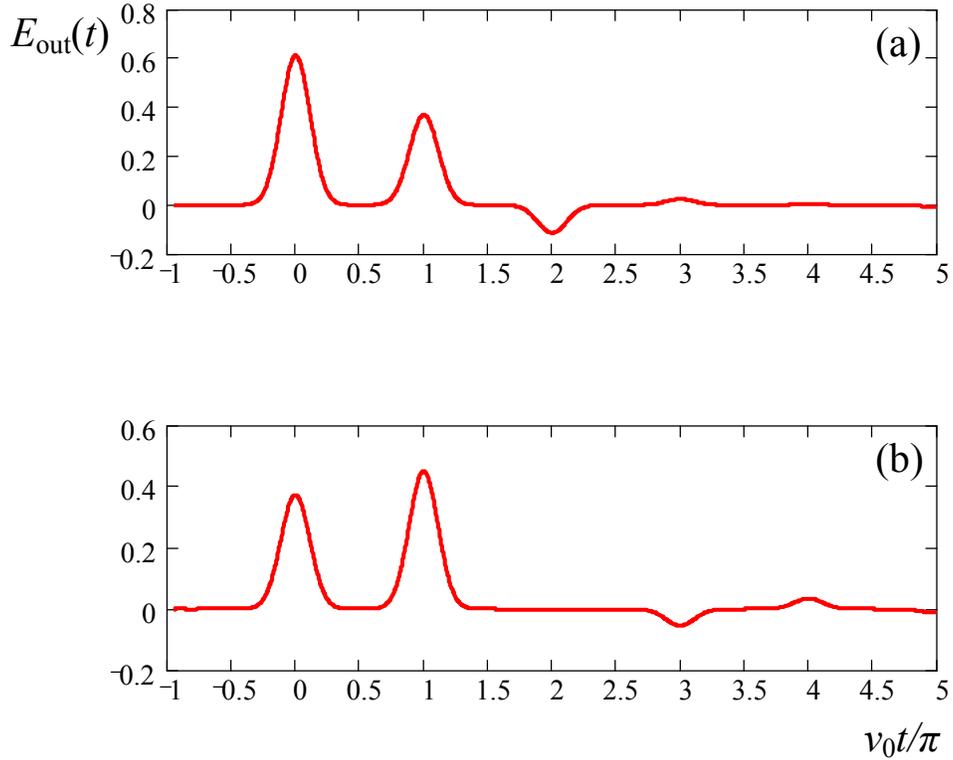}}\caption{Time
dependence of the field $E_{\text{out}}(t)$ (normalized to $E_{0}$) at the
exit of the medium for the AHFC II with $d_{p}=4$ (a) and 8 (b).}%
\label{fig:10}%
\end{figure}\begin{figure}[ptbptbptb]
\resizebox{0.5\textwidth}{!}{\includegraphics{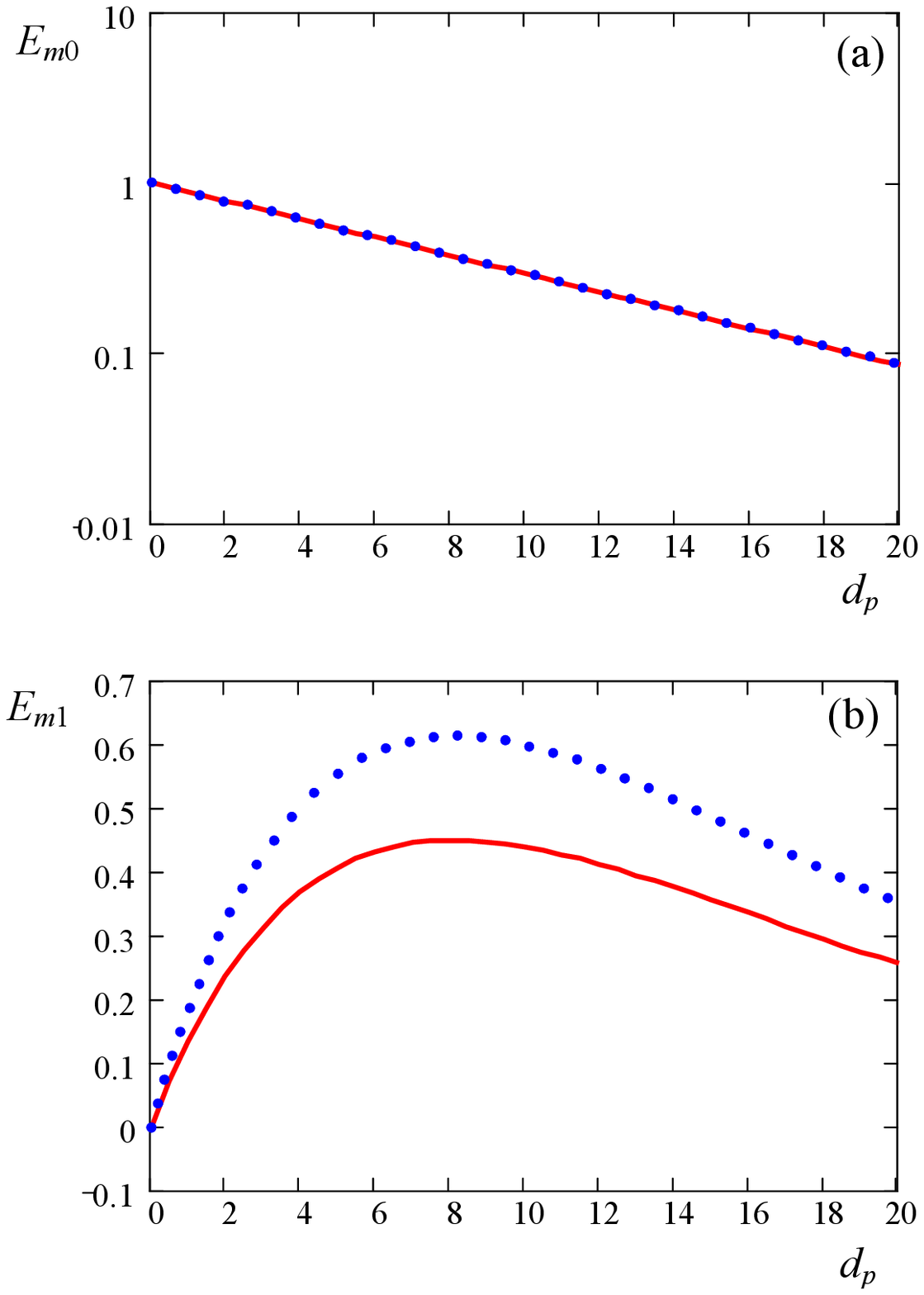}}\caption{ Dependences
of the maximum amplitude of the pulse with no delay(a) and the first pulse
with delay $t_{d1}=\pi/\nu_{0}$ (b) on optical thickness $d_{p}$. Solid line
in red corresponds to AHFC II and dotted line in blue to AHFC I.}%
\label{fig:11}%
\end{figure}Time dependence of the field $E_{\text{out}}(t)$ at the exit of
the medium for the AHFC II is shown in Fig. 10. It is calculated numerically
for the Gaussian input pulse with the help of the equation%
\begin{equation}
E_{out}(t)=\frac{1}{2\pi}\int_{-\infty}^{+\infty}E_{in}(\nu)e^{-i\nu
t-T_{L}(\nu)}dt. \label{Eq63}%
\end{equation}
Analysis of the numerical results shows that maximum amplitude of the pulse
with no delay, $E_{m0}$, taking place at $t=0$, has the same dependence on the
optical thickness $d_{p}$, as for the AHFC I, see Fig. 11 (a). This
contradicts the finesse concept of the frequency combs, which predicts the
amplitude $E_{m0}=E_{0}\exp(-d_{p}/2F)$, where $F$ is the comb finesse. For
the AHFC I the value $F$ was estimated as $F_{\mathrm{exc}}=4.063$ (see Sec.
VII), while for AHFC II its finesse can be estimated as $\nu_{0}/\Gamma
_{0}=6.37$, which is $1.5$ times larger than that following from the
dependence $E_{m0}$ on the optical thickness $d_{p}$, shown in Fig 11 (a) by
solid line in red. This difference could be explained by nonzero absorption at
the center of the transmission windows, see Fig. 9. Therefore, the pulse with
no delay reduces more than it is predicted by the exponent $\exp(-d_{p}/2F)$.
The effect of remnant absorption at the centers of the transmission windows
was taken into account in Ref. \cite{Gisin2008,Gisin2009}.

Maximum amplitude of the first delayed pulse, $E_{m1}$, is obtained for both,
AHFC I and AHFC II, at the same value of the optical thickness, $d_{p}\simeq
8$, see Fig. 11 (b). However, for AHFC II this maximum is smaller.

To estimate the amplitudes $E_{m0}$, $E_{m1}$\ of the first two pulses with no
delay ($t_{0}$) and delayed ($t_{1}$) we express the transmission function as
follows%
\begin{equation}
T_{L}(\nu)=\frac{d_{p}}{2}\sum_{k=0}^{+\infty}a_{k}e^{ik\pi\nu/\nu_{L}},
\label{Eq64}%
\end{equation}
where the coefficients in the sum are%
\begin{equation}
a_{k}=\frac{1}{2\nu_{0}}\int_{-\nu_{0}}^{\nu_{0}}\frac{2}{d_{p}}T_{L}%
(\nu)e^{-ik\pi\nu/\nu_{L}}d\nu. \label{Eq65}%
\end{equation}
Then, the amplitude of the radiation field at the exit of the medium is
described by equation%
\begin{equation}
E_{\text{out}}(t)=e^{-a_{0}d_{p}/2}\left[  E_{\text{in}}\left(  t\right)
-B_{1}(d_{p})E_{\text{in}}\left(  t-\frac{\pi}{\nu_{L}}\right)  -B_{2}%
(d_{p})E_{\text{in}}\left(  t-\frac{2\pi}{\nu_{L}}\right)  +...\right]  ,
\label{Eq66}%
\end{equation}
where%
\begin{equation}
B_{1}(d_{p})=a_{1}\frac{d_{p}}{2}, \label{Eq67}%
\end{equation}
and%
\begin{equation}
B_{2}(d_{p})=a_{2}\frac{d_{p}}{2}-a_{1}^{2}\frac{d_{p}^{2}}{8}. \label{Eq68}%
\end{equation}
The coefficients $a_{k}$ can be easily calculated. For example, for the
infinite sum of Lorentzians [infinite sum in Eq. (\ref{Eq59})] the
coefficients are%
\begin{equation}
a_{0}=\frac{\pi\Gamma_{0}}{2\nu_{L}}, \label{Eq69}%
\end{equation}
and%
\begin{equation}
a_{k}=(-1)^{k}\frac{\pi\Gamma_{0}}{\nu_{L}}e^{-k\pi\Gamma_{0}/\nu_{L}},
\label{Eq70}%
\end{equation}
where $k=1,2,..$.

Maximum amplitude of the first delayed pulse is achieved when $d_{p}=2/a_{0}$.
For this value of the optical thickness $d_{p}$ we have%
\begin{equation}
-e^{-a_{0}d_{p}/2}B_{1}(d_{p})=2e^{-1-\pi\Gamma_{0}/\nu_{L}}. \label{Eq71}%
\end{equation}
Thus, the amplitude of the first pulse with a delay $t_{1}$ cannot be larger
than $73.6\%$ of the amplitude of the input pulse since $2e^{-1}=0.736$. This
maximum amplitude is achieved if the condition $\pi\Gamma_{0}/\nu_{L}\ll1$ is
satisfied and $d_{p}=4\nu_{L}/\pi\Gamma_{0}\gg1$.

In the numerical example, shown in Fig. 10, the parameter characterizing the
absorption peaks is $\Gamma_{0}=0.157\nu_{0}$, where $\nu_{0}=\nu_{L}$. For
this relation between $\Gamma_{0}$ and $\nu_{0}$\ one can calculate numerical
values of $a_{0}$, $a_{1}$, and $a_{2}$ with the help of Eqs. (\ref{Eq65}),
(\ref{Eq69}), and (\ref{Eq70}). They are $a_{0}=0.246$, $a_{1}=-0.3$,
$a_{2}=0.184$ and describe excellent the amplitudes of the first three pulses
at the exit of the absorber. The coefficient $a_{0}=0.246$ gives the exact
value of the finesse for this comb $F=a_{0}^{-1}$, which coincides with the
numerically found value $4.063$. Maximum amplitude of the first delayed pulse
is achieved when $d_{p}=2/a_{0}=8.13$. For this value of optical thickness the
amplitude $E_{m1}$ is estimated as $E_{0}a_{1}/(a_{0}e)=0.449$.

In Ref. \cite{Bonarota2012} it was shown that pulses with number $k$,
generated at the exit of the medium with the periodic structure of Lorentzian
absorption peaks in the spectrum, delay slightly with respect time $t_{k}$
when they have to appear. I suppose that this delay is caused by imperfect
shape of the sequence of the absorption peaks. This sequence is created,
first, by burning a broad hole with a rectangular shape and then, by bringing
atoms back in particular positions within the hole building a periodic
structure of the absorption peaks. However, the broad hole could have a shape
of the dip with smooth edges and the heights of the peaks, created afterwards,
could form a smooth bell-shaped envelope. Such a structure with nonconstant
heights of the peaks and nonuniform depths of the transparency windows could
be responsible for the observed delay of the generated pulses.

\section{Conclusion}

The propagation of light pulse in a medium with a periodic structure in the
absorption spectrum is analyzed. Two periodic structures, harmonic and
anharmonic, are considered. Both are idealized as having infinite spectrum
consisting of absorption peaks separated by transparency windows. Frequency
dependent complex dielectric constants are derived for these periodic
structures with the help of Kramers-Kronig relation and solution of the Bloch
equations. The method of solution of Maxwell-Bloch equation describing the
pulse propagation in a frequency periodic medium (frequency crystal) is
proposed. For the harmonic frequency crystal the exact solution is obtained in
a simple form. For the anharmonic frequency crystal simple analytical solution
describing the first three pulses in the pulse sequence at the exit of the
crystal is derived. Filtering a short pulse through the harmonic frequency
crystal allows to generate a pulse sequence separated by long time intervals.
These pulses are coherent and could be applied to create time-bin qubits since
the phases and amplitudes of the pulses can be manipulated by adjusting the
parameters of the frequency crystal. HFC and AHFC can be created in the
inhomogeneously broadened absorption spectrum of crystals with
rear-earth-metal impurity ions by the hole burning technique or by
constructing the system of many mini or micro-cavities interacting with a
common broadband cavity coupled with external waveguide.

\section{Acknowledgements}

The author expresses his thanks to Prof. Tagirov for useful discussions and
help in the manuscript preparation. This work was partially funded by the
Program of Russian Academy of Sciences "Actual problems of the low temperature
physics," the Program of Competitive Growth of Kazan Federal University funded
by the Russian Government.


\begin{thebibliography}{99}                                                                                               %


\bibitem {Wilczek}F. Wilczek, Phys. Rev. Lett. \textbf{109}, 160401 (2012).

\bibitem {Sacha}K. Sacha, Phys. Rev. A \textbf{91}, 033617 (2015).

\bibitem {Khemani}V. Khemani, A. Lazarides, R. Moessner, and S. L. Sondhi,
Phys. Rev. Lett. \textbf{116}, 250401 (2016).

\bibitem {Else}D. V. Else, B. Bauer, and C. Nayak, Phys. Rev. Lett.
\textbf{117}, 090402 (2016).

\bibitem {Sondhi}C. W. von Keyserlingk, V. Khemani, and S. L. Sondhi, Phys.
Rev. B \textbf{94}, 085112 (2016).

\bibitem {Yao}N. Y.Yao, A. C. Potter, I.-D. Potirniche, and A. Vishwanath,
Phys. Rev. Lett. \textbf{118}, 030401 (2017).

\bibitem {Lukin}S. Choi et al., \textbf{543}, 221 (2017).

\bibitem {Gisin2008}H. de Riedmatten, M. Afzelius, M. U. Staudt, C. Simon, and
N. Gisin, Nature \textbf{456}, 773 (2008).

\bibitem {Gisin2009}M. Afzelius, C. Simon, H. de Riedmatten, and N. Gisin,
Phys. Rev. A \textbf{79}, 052329 (2009).

\bibitem {Lauro}R. Lauro, T. Chaneli\`{e}re, and J.-L. Le Gou\"{e}t, Phys.
Rev. A \textbf{79}, 053801 (2009).

\bibitem {Bonarota}M. Bonarota, J. Ruggiero, J.-L. Le Gou\"{e}t, and T.
Chaneli\`{e}re, Phys. Rev. A \textbf{81}, 033803 (2010).

\bibitem {Chaneliere}T. Chaneli\`{e}re, J. Ruggiero, M. Bonarota, M. Afzelius,
and J.-L. Le Gou\"{e}t, New Journal of Physics \textbf{12}, 023025 (2010).

\bibitem {Bonarota2012}M. Bonarota, J.-L. Le Gou\"{e}t, S. A. Moiseev, and T.
Chaneli\`{e}re, J. Phys. B: At. Mol. Opt. Phys. \textbf{45}, 124002 (2012).

\bibitem {Vagizov}F. Vagizov, V. Antonov, Y. V. Radeonychev, R. N.
Shakhmuratov, and O. Kocharovskaya, Nature \textbf{508}, 80 (2014).

\bibitem {Shakhmuratov15}R. N. Shakhmuratov, F. G. Vagizov, V. A. Antonov, Y.
V. Radeonychev, M. O. Scully, and Olga Kocharovskaya, Phys. Rev. A
\textbf{92}, 023836 (2015).

\bibitem {Shakhmuratov17}R. N. Shakhmuratov, Phys. Rev. A \textbf{95}, 033805 (2017).

\bibitem {McAuslan}D. L. McAuslan, L. R. Taylor, and J. J. Longdella. Appl.
Phys. Lett. \textbf{101}, 191112 (2012).

\bibitem {Horiuchi}N. Horiuchi, Nature Photonics \textbf{7}, 85 (2013).

\bibitem {Crisp}M. D. Crisp, Phys. Rev. A 1, 1604 (1970).

\bibitem {Abramowitz}\textit{Handbook of Mathematical Functions}, edited by M.
Abramowitz and I. A. Stegun (Dover, New York, 1965).

\bibitem {Shakhmuratov1}R. N. Shakhmuratov and J. Odeurs, Phys. Rev. A
\textbf{71}, 013819 (2005).

\bibitem {Shakhmuratov2}R. N. Shakhmuratov, A. Rebane, P. Megret, and J.
Odeurs, Phys. Rev. A \textbf{71}, 053811 (2005).

\bibitem {Moiseev}S. A. Moiseev, K. I. Gerasimov, R. R. Latypov, N. S.
Perminov, K. V. Petrovnin, and O. N. Sherstyukov, Scientific Reports
\textbf{8}, 3982 (2018).
\end{thebibliography}
\end{document}